\documentclass[conference]{IEEEtran}

\usepackage{graphicx}
\usepackage[english]{babel}
\usepackage{amssymb}
\usepackage{amsbsy}
\usepackage{verbatim}
\usepackage[cmex10]{amsmath}
\usepackage[varg]{txfonts}
\let\mathbb=\varmathbb
\DeclareSymbolFont{letters}{OML}{ztmcm}{m}{it}
\usepackage{tikz}
\usepackage{pgfplots}
\usepackage{multirow}
\usepackage{array}
\usepackage{cite}

\newtheorem{definition}{Definition}
\usepackage[noend,ruled]{algorithm2e}

\usepackage{tikz-timing}

\newcommand{\fixme}[2]{\ifx&#2&{\color{red}#1}\else{\color{red}FIXME\{}#1{\color{red}\}}\footnote{{\color{red}#2}}\PackageWarning{Fixme}{#1: #2}\fi}

\usepackage{ifpdf}
\ifpdf
\title{Stall Pattern Avoidance in \\Polynomial Product Codes}

\author{Carlo Condo, Fran\c{c}ois Leduc-Primeau, Gabi Sarkis, Pascal Giard, \emph{Member, IEEE},\\
and Warren J. Gross, \emph{Senior Member, IEEE}\\
Department of Electrical and Computer Engineering, McGill University, Montreal, QC, Canada}

\begin{document}
\maketitle
\begin{abstract}
Product codes are a concatenated error-correction scheme that has been often considered for applications requiring very low bit-error rates, which demand that the error floor be decreased as much as possible. In this work, we consider product codes constructed from polynomial algebraic codes, and propose a novel low-complexity post-processing technique that is able to improve the error-correction performance by orders of magnitude. We provide lower bounds for the error rate achievable under post processing, and present simulation results indicating that these bounds are tight.
\end{abstract}

% Usually, IEEE conference papers shouldn't feature keywords
% \begin{IEEEkeywords}
% Product codes, optical communications, error floor, post processing
% \end{IEEEkeywords}

\section{Introduction}
\label{sec:intro}

Product codes \cite{Elias_IRE54} are concatenated codes often considered for applications requiring high throughput and very low bit error rate (BER) \cite{Jian_GLOBECOM13,LeBidan_EURASIP08}.
They allow efficient construction of long codes from short component codes. The concatenated structure and code length guarantee very good error-correction performance, while low decoding latency and high throughput can be achieved by choosing simple component codes and by exploiting the inherent parallelism of product codes.

Polynomial algebraic codes like Bose-Chaudhuri-Hocquenghem (BCH) codes \cite{Bose_IC60} and Reed-Solomon (RS) codes \cite{Reed_JSIAM60} are able to detect $t$ errors and identify their position in the codeword.
They can be decoded efficiently with hard-decision algorithms; moreover, these algorithm can undergo substantial speed-up and complexity reduction when applied to particular codes. They have been considered in the past as component codes for product codes.

Forward error correction (FEC) schemes can incur an error-correction performance degradation at low BER. This sudden decrement of the BER curve slope is known as an error floor, which is usually caused by particular error patterns that are difficult for the decoder to correct. For a hard-decision product-code decoder, these patterns are often called \emph{stall} patterns \cite{smith:2012} or stopping sets.

To avoid the occurrence of an error floor, and thus meet extremely low BER requirements, post processing can be employed \cite{Emmadi_PP15}. In this paper, we introduce a novel post-processing technique that dramatically increases the error-correction performance of polynomial product codes. Simulations show that the introduction of post-processing is able to lower the BER by three orders of magnitude.

The remainder of the paper is organized as follows. Section~\ref{sec:prod} introduces product-code encoding and decoding. Section~\ref{sec:stall} analyzes stall patterns and their contribution to the error floor. The proposed post-processing technique is detailed in Section~\ref{sec:PP}, and its impact on the error-correction performance is evaluated in Section~\ref{sec:perf}. Conclusions are drawn in Section~\ref{sec:conc}.

\section{Product Codes}
\label{sec:prod}

%Description of the FEC scheme, post-processing included.

Product codes are a class of error-correcting codes constructed through parallel concatenation. They were introduced in \cite{Elias_IRE54} and generalized in \cite{Tanner_TIT81}. The encoding process starts by forming a matrix of information symbols; then, the rows of the matrix are encoded using the row component code. Afterwards, the columns of the matrix are encoded using the column component code. FEC schemes based on code concatenation \cite{Tychopoulos_CDM06,Lee_ISOCC10,Justesen_TCOM11,Chen_ICCT12,Jian_GLOBECOM13} have been widely adopted in the past in applications targeting very low BER. These schemes intertwine simpler codes greatly enhancing their error-correction performance. Reliable and well studied algebraic codes like BCH and RS codes have been used as component codes for product codes for decades.

Let us define a generic polynomial algebraic code ($n$,~$k$,~$t$), where $n$ is the code length in symbols, $k$ the number of information symbols, and $t$ the number of errors that the code can detect and correct. %Figure~\ref{fig:prodEnc} shows the structure of 
Given a generic matrix of $k_2 \times k_1$ symbols, a product code matrix can be obtained by encoding the rows with ($n_1$,~$k_1$,~$t_1$), and the columns with ($n_2$,~$k_2$,~$t_2$). For simplicity, in the remainder of the paper we assume that both rows and columns have been encoded with the same component code ($n$,~$k$,~$t$): the generalization of the presented results is straightforward.
	
Product-code iterative hard-decision decoding was first described in \cite{Abramson_TCT68}. The decoding process follows the same schedule as product-code encoding: first, the rows are decoded with the row component decoder, and then the columns are decoded with the column component decoder. The process is repeated for a set number of iterations. The decoding latency depends on the number of iterations and on the complexity of the component decoding. Moreover, product codes allow for a high degree of parallelism in the decoder, since all rows (columns) of the product-code matrix can be decoded concurrently: consequently, the hardware architecture of the decoder plays an important role in determining the decoding latency.

On the other hand, the error-correction performance of a product code mainly depends on the choice of the component codes, together with their length and rate. Increasing the number of decoding iterations can substantially improve the error-correction performance at high to medium BER. However, this technique will prove mostly ineffective at very low BER.

% \begin{figure}[tbp]
%   \begin{center}
%     \includegraphics[scale=0.6]{./Figures/prodMatr.pdf}
%     \caption{Product-code matrix.}
%     \label{fig:prodEnc}
%   \end{center}
% \end{figure}	

\section{Error Floor and Stall Pattern Analysis} \label{sec:stall}

Error-correcting codes can incur a flattening of the BER curve at low BER, that cannot be overcome by improving the channel conditions or increasing the number of decoding iterations. This degradation of the error-correction performance is called an error floor.
An error floor is usually caused by combinations of errors that are hard to detect and correct. 
In the specific case of product codes decoded with hard-decision algorithms, these are known as stall patterns \cite{smith:2012}. Since polynomial codes are able to detect and correct $t$ errors, stall patterns are defined as follows.
%Let $r_i$ be the set of codeword positions corresponding to row $i$, and $c_j$ be the set of codeword positions corresponding to column $j$.
Let us consider a generic product code matrix. Let $r_i$ be the set of symbols in the row component codeword $i$, and $c_j$ be the symbols in the column component codeword $j$.
\begin{definition}
	A stall pattern is a set $S$ of codeword symbols with the following properties:
  \begin{enumerate}
  \item If $s_{i,j} \in S$ then $|r_i \cap S| > t$, $\forall (i,j)$,
  \item If $s_{i,j} \in S$ then $| c_j \cap S | > t$, $\forall (i,j)$,
  \end{enumerate}
  where $s_{i,j}$ is the symbol at the intersection of row $i$ and column $j$.
\end{definition}
	
To perform our analysis, we pessimistically assume that decoding fails if a stall pattern exists in the received frame, that is we neglect the possibility that undetected errors in the component decoders cause the product decoder to avoid the stall \cite{smith:2012}. 
To obtain a lower bound, we can consider the probability that a minimal stall pattern occurs in the received frame, i.e. error patterns $S$ for which $s_{i,j}\in S \Rightarrow |r_i \cap S|=t+1$ and $|c_j \cap S|=t+1$.
We expect the error floor to be dominated by stall patterns present in the received frame \cite{smith:2012,Justesen_TCOM11}, and the probability of occurrence of a stall pattern to be inversely proportional to its cardinality. Therefore such a lower bound should be reasonably tight.

To count minimal stall patterns, we first select the $t+1$ rows that will be affected, and then select the $t+1$ column positions that will be shared by all the affected rows. The number $M$ of minimal patterns is thus given by
\begin{equation}
M = {n \choose t+1}^2.
\end{equation}
We consider bit errors to be independent: thus, for a binary code, the probability that a minimal stall pattern occurs in the received vector at specific bit positions is $p^{(t+1)^2}$, where $p$ is the channel BER. The probability of observing any minimal stall pattern in the received vector is then $M p^{(t+1)^2}$. When a decoding failure due to a minimal stall occurs, $(t+1)^2$ of the $n^2$ bits in the frame are in error, and therefore the BER is lower bounded by
\begin{equation}\label{eq:minStallFloor}
f_\mathrm{min}(p)= \frac{(t+1)^2 M p^{(t+1)^2}}{n^2}. %Note: This expression matches (7) in \cite{justesen:2011}
\end{equation}
If instead the component codes are non-binary codes able to correct up to $t$ symbols of $b$ bits, 
we can obtain a lower bound on the Symbol Error Rate (SER) by taking $f_\mathrm{min}(p_s)$, where $p_s$ is the channel SER and is given by $p_s=1-(1-p)^b$ for binary memoryless channels. 
For the non-binary case, the bound $g_\mathrm{min}(p)$ on the BER can be expressed as
\begin{equation} \label{eq:minStallFloorNonbin}
g_\mathrm{min}(p)= \frac{(b-1)p+1}{b^2} f_\mathrm{min}\left(1-(1-p)^b\right),
% \frac{((b-1)p+1) (t+1)^2 M p_s^{(t+1)^2}}{n^2b^2} 
\end{equation}
where $(b-1)p+1$ is the expected number of wrong bits in an erroneous symbol.

Stall-pattern avoidance can greatly benefit from the use of \emph{extended}-polynomial codes. 
An extended binary polynomial code of length $n+1$ is composed of a binary polynomial code of length $n$ and of an additional parity bit. While this additional parity bit does not improve the number of errors $t$ that can be corrected by the code, it increases to $t+1$ the errors that can be detected, with a small cost in terms of code rate. Consequently, a product code based on extended-polynomial codes can detect decoding failures caused by minimal stall patterns. Non-binary polynomial codes like RS can be extended by adding a $b$-bit parity symbol instead of a single parity bit. The parity symbol is the sum of all the codeword symbols performed over the Galois Field of order $b$. %, \fixme{i.e. the XOR operation}{}.
This means that the $i^{th}$ bit of the parity symbol can be computed by calculating the parity of the $i^{th}$ bit of every codeword symbol. As long as the component polynomial code is linear, every row and every column of the product code based on its extended version is a valid component codeword.

Algorithm \ref{alg:extended} portrays how the additional parity bit can be used in the decoding of binary polynomial component codes. Without loss of generality, we assume that this parity bit is placed at position $n+1$ in the codeword, and we identify it as $r_{n+1}$. The $\mathrm{CD}$ function refers to the standard component decoder, which returns a flag $\textsc{fail}$ indicating whether or not the decoder detected a failure, and in case it succeeded a vector $e$ of length $n$ indicating the location of errors. The notation $x_{i:j}$ with $i\leq j$ refers to a vector of length $j-i+1$ containing elements $i,i+1,\dots,j$ of the vector $x$. The operator $\oplus$ denotes modulo-2 addition. The extended-code decoder declares failure either in case $\textsc{fail}$ is risen or in case $d=t$ and $pr\ne 0$, i.e. when $t$ errors have been detected but the parity check has failed.

\begin{algorithm}[tb] 
  \SetKwInOut{Input}{input} \SetKwInOut{Output}{output}
  \DontPrintSemicolon

  \Input{Component codeword $r$}
  \Output{Updated codeword $r'$}
  
  \Begin{
    $\textsc{fail}, e \gets \mathrm{CD}(r_{1:n})$ \;
		\If{$\textsc{fail}$}{
			$r' \gets r$\;
		}
		\Else{
			$d:= \sum_{i=1}^{n} e_i$ \;
      $pr:= \left(d + \sum_{i=1}^{n+1} r_i\right) \mod 2$ \;
      
      \If{$d < t$}{
        $r'_{1:n} \gets r_{1:n} \oplus e$ \;
        $r'_{n+1} \gets r_{n+1} \oplus p$ \;
      }
      \ElseIf{$pr=0$}{
        $r'_{1:n} \gets r_{1:n} \oplus e$ \;
        $r'_{n+1} \gets r_{n+1}$ \;
      }
			\Else{
				$r' \gets r$\;
			}
		}
  }
  \caption{Decoding of extended binary polynomial codes.}
  \label{alg:extended}
\end{algorithm}

Algorithm \ref{alg:RSext} describes the decoding process in case of extended non-binary polynomial codes, where error vector $e^{sym}$ identifies the wrong symbols. % \fixme{It differs from Algorithm \ref{alg:extended} in that failure is declared if the decoder rises the $\textsc{fail}$ signal or if it detects $t$ erroneous symbols and the received parity symbol mismatches the computed one.}{}

\begin{algorithm}[tb] 
  \SetKwInOut{Input}{input} \SetKwInOut{Output}{output}
  \DontPrintSemicolon

  \Input{Component codeword $r$}
  \Output{Updated codeword $r'$}
  
  \Begin{
    $\textsc{fail}, e^{sym}, e \gets \mathrm{CD}(r_{1:n})$ \;
		\If{$\textsc{fail}$}{
			$r' \gets r$\;
		}
		\Else{
		$d:= \sum_{i=1}^{n} e^{sym}_{i}$ \;
		\For{$w=1:b$}{
		$d^b_w:= \sum_{i=0}^{n-1} e_{ib+w}$ \;
      $pr_w:= \left(d^b_w + \sum_{i=0}^{n} r_{ib+w}\right) \mod 2$ \;
      }
      \If{$d < t$}{
        $r'_{1:nb} \gets r_{1:n} \oplus e$ \;
        $r'_{nb+1:nb+b} \gets r_{nb+1:nb+b} \oplus pr_{1:b}$ \;
      }
      \ElseIf{$pr_{1:b}=0$}{
        $r'_{1:nb} \gets r_{1:nb} \oplus e$ \;
        $r'_{nb+1:nb+b} \gets r_{nb+1:nb+b}$ \;
      }
			\Else{
				$r' \gets r$\;
			}
		}
  }
  \caption{Decoding of extended non-binary polynomial codes.}
  \label{alg:RSext}
\end{algorithm}

\section{Stall Pattern Post Processing}
\label{sec:PP}

The post-processing technique mentioned in \cite{Emmadi_PP15} is applied to the binary erasure channel: all symbols of the row component codes whose decoding has failed are changed to erasures, then column component codes are decoded. Failed columns are changed to erasures as well, and the process is repeated until no failures are detected.
We take a different approach to post processing. The structure of a product code guarantees that after a full decoding iteration (rows + columns), any bit in error is located at the intersection of a row codeword and of a column codeword that are both in error. Some error locations can thus be identified from the extended-code row and column decoding failures. Note however that bit positions located at the intersection of a row and of a column that are both in error are not necessarily in error. As shown below, newly introduced errors can be corrected by an additional decoding iteration.

We propose a post-processing algorithm that inverts the bits located at the intersection of rows and columns that are known to be in error. It is performed after a set number of product-code decoding iterations have been completed.
Let us denote by $R$ the set of row indices for which the extended-component decoder reported a decoding failure. Similarly we denote by $C$ the set of column indices for which the extended-code decoding has failed. 
\begin{itemize}
\item \emph{Binary component codes}: if $|R|>0$ and $|C|>0$, we flip the bit located at the intersection of each row in $R$ with each column in $C$. Since this may introduce new bit errors, we then perform one full product decoding iteration.
\item \emph{Non-binary component codes}: in this case, the intersection of a row and a column is a $b$-bit symbol. However, it is possible to identify which bits of the symbol are most probably wrong by using the parity symbols. Let us take row $i\in R$ and column $j\in C$, and recompute their respective parity symbols. A bit $w$ in symbol ($i$,$j$) will be flipped if bit $w$ of the row or column parity symbol is different from the corresponding bit in the recomputed parity symbol. 
For example, if the row $i$ parity symbol has mismatching bits $1$ and $3$, and column $j$ parity symbol has mismatching bits $3$ and $4$, we flip bits $1$, $3$ and $4$ in symbol ($i$,$j$).
If instead neither row $i$ nor column $j$ have mismatching bits in their parity symbols, we flip all bits of symbol ($i$,$j$). Like with binary component codes, a full product decoding iteration follows the bit flipping.

%and suppose that the $w$-th and $x$-th bit of the parity symbol of row $i$ mismatch the computed parities, as does the $\ell$-th bit in the parity symbol of column $j$. We then flip bit $w$, bit $x$ and bit $\ell$ in symbol ($i$,$j$). 

\end{itemize}

\begin{table}[tp]
	\caption{Stall pattern multiplicity}
	\vspace{0.1cm}
	\begin{center}
    %	\scalebox{1}{
    \begin{tabular}{cc}
      \hline
      $n_e$ & $m(n_e)$\\
      \hline
      12 & 8\\
      14 & 72\\
      15 & 16\\
      16 & 1\\
      \hline
    \end{tabular}
    % }
	\end{center}	
	\label{tbl:stallmult}
	\vspace{-10pt}
\end{table}%
	
Since the codeword positions involved in the post processing are not guaranteed to be in error, it is non-trivial to determine which error patterns can indeed be resolved using this algorithm.
It can correct minimal stall patterns, since all bits in error correspond to error-row and error-column intersections. On the other hand, stall patterns $S$ such that $s_{i,j} \in S \Rightarrow |r_i \cap S|=t+2$ and $|c_j \cap S|=t+2$ are clearly not correctable, since none of the incorrect rows or columns can be detected. %There might be other uncorrectable patterns among the patterns that can be described by a $(t+2) \times (t+2)$ matrix.
	
For small values of $t$, it is possible to perform an exhaustive pattern search to determine which ones are correctable using post processing and which ones are not. Let $n_e = |S|$ be the cardinality of a stall pattern $S$, and let $m(n_e)$ be the number of patterns with cardinality $n_e$ such that $s_{i,j} \in S \Rightarrow t<|r_i \cap S|\leq t+2$ and $t<|c_j \cap S|\leq t+2$.
The BER of a binary code is then lower bounded by
	\begin{equation} \label{eq:contrib}
	f_\mathrm{pp}(p)= \frac{M}{(n+1)^2} \sum_{n_e} m(n_e) \cdot p^{n_e} \cdot (1-p)^{(t+2)^2-n_e} \cdot n_e,
	\end{equation}
	where 
	\begin{equation}\label{eq:M}
	M = {n+1 \choose t+2}^2.
	\end{equation}
As for the analysis presented in Section~\ref{sec:stall}, in the case of non-binary codes, \eqref{eq:contrib} becomes a lower bound on the SER if the channel BER $p$ is replaced with the channel SER $p_s=1-(1-p)^b$. The BER lower bound for non-binary codes is instead expressed by
\begin{equation} \label{eq:contribNonbin}
g_\mathrm{pp}(p)= \frac{(b-1)p+1}{b^2} f_\mathrm{pp}\left(1-(1-p)^b\right).
%  ((b-1)p+1)\frac{M}{(n+1)^2b^2} \sum_{n_e} m(n_e) \cdot p_s^{n_e} \cdot (1-p_s)^{(t+2)^2-n_e} \cdot n_e.
\end{equation}
For $t=2$, the non-zero values of $m(n_e)$ obtained using an exhaustive search are listed in Table~\ref{tbl:stallmult}.

%The BER curve obtained with post processing is shown to approach its error floor with an greater steepness than the BER curve without post processing. This is because at the 
%\fixme{Comment on why the two-iteration curve is not as close as the four-iteration curve to the floor?}{}

\section{Error-Correction Performance}
\label{sec:perf}

\begin{figure}[tbp]
  \begin{center}
    \includegraphics[scale=0.41]{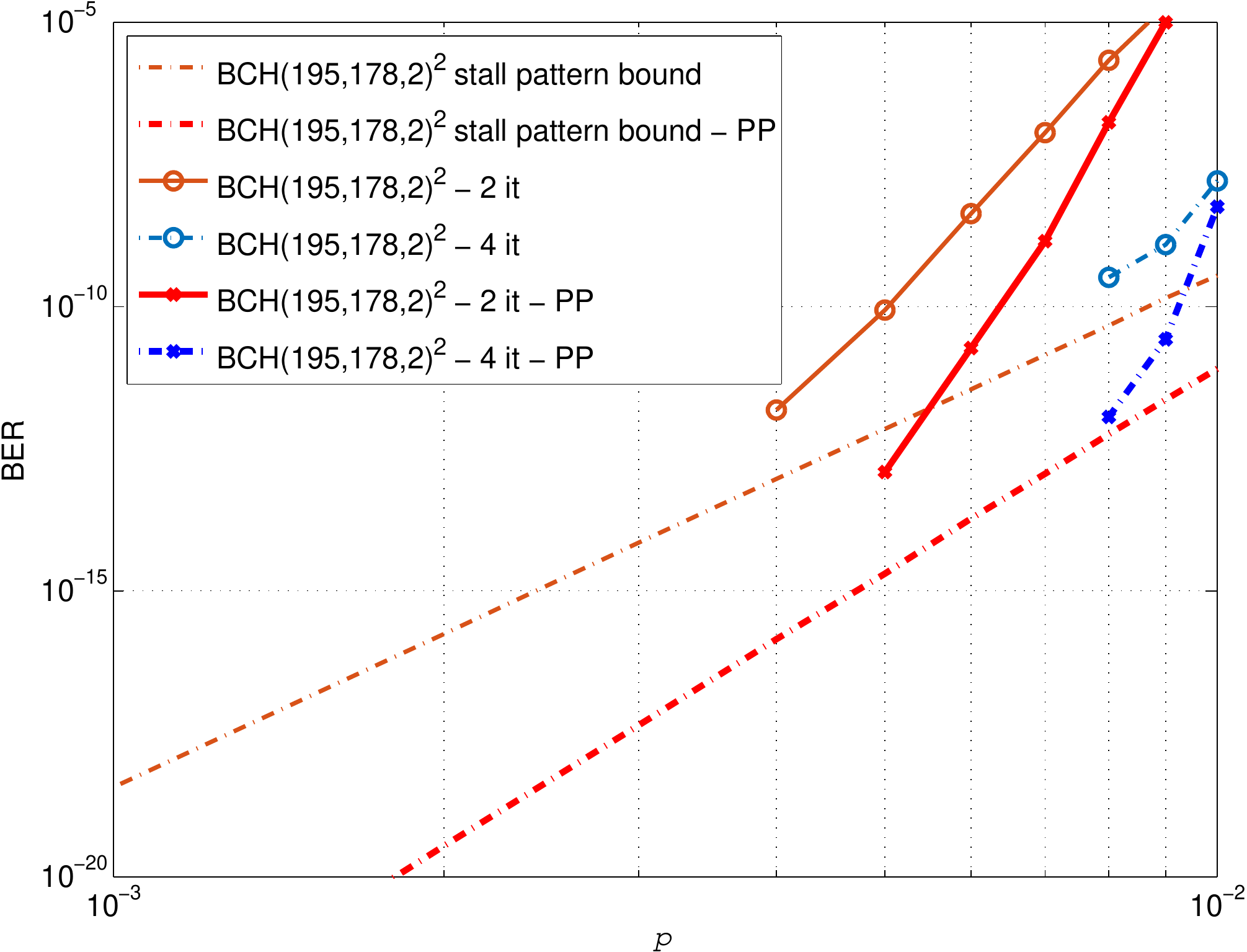}
    \caption{Stall-pattern contribution bounds and BER curves for extended BCH-based (195,178,2)$^2$ product code.}
    \label{fig:BCHbounds}
  \end{center}
\end{figure}

Figure~\ref{fig:BCHbounds} shows the BER and the stall-pattern contribution to the error floor for the product code based on the (195,~178,~2) extended BCH code, with and without post processing, for two and four decoding iterations. It can be seen that the error-correction performance of the code is greatly enhanced for all values of $p$, and that the error bound is lowered and increased in steepness by post processing. At $p=3\times10^{-3}$, the contribution of stall patterns to the error floor is decreased by more than three orders of magnitude. At the same time, post processing allows substantial BER gain under all decoding conditions. The tightness of the bound can be noticed in case of four decoding iterations and post processing, where the error floor is reached at higher $p$ values.

%Figure~\ref{fig:stallcontrib} shows the BER and the stall pattern contribution to the error floor for the product codes based on the (195,~178,~2) extended BCH code and the (32,~27,~2) extended RS code, with and without post processing, for two decoding iterations. It can be seen that the error-correction performance of the code is greatly enhanced for all values of $p$, and that the error bound is lowered and increased in steepness by post processing. At $p=3\times10^{-3}$, the contribution of stall patterns to the error floor is decreased by more than three orders of magnitude.

\begin{figure}[tbp]
	\begin{center}
    \includegraphics[scale=0.44]{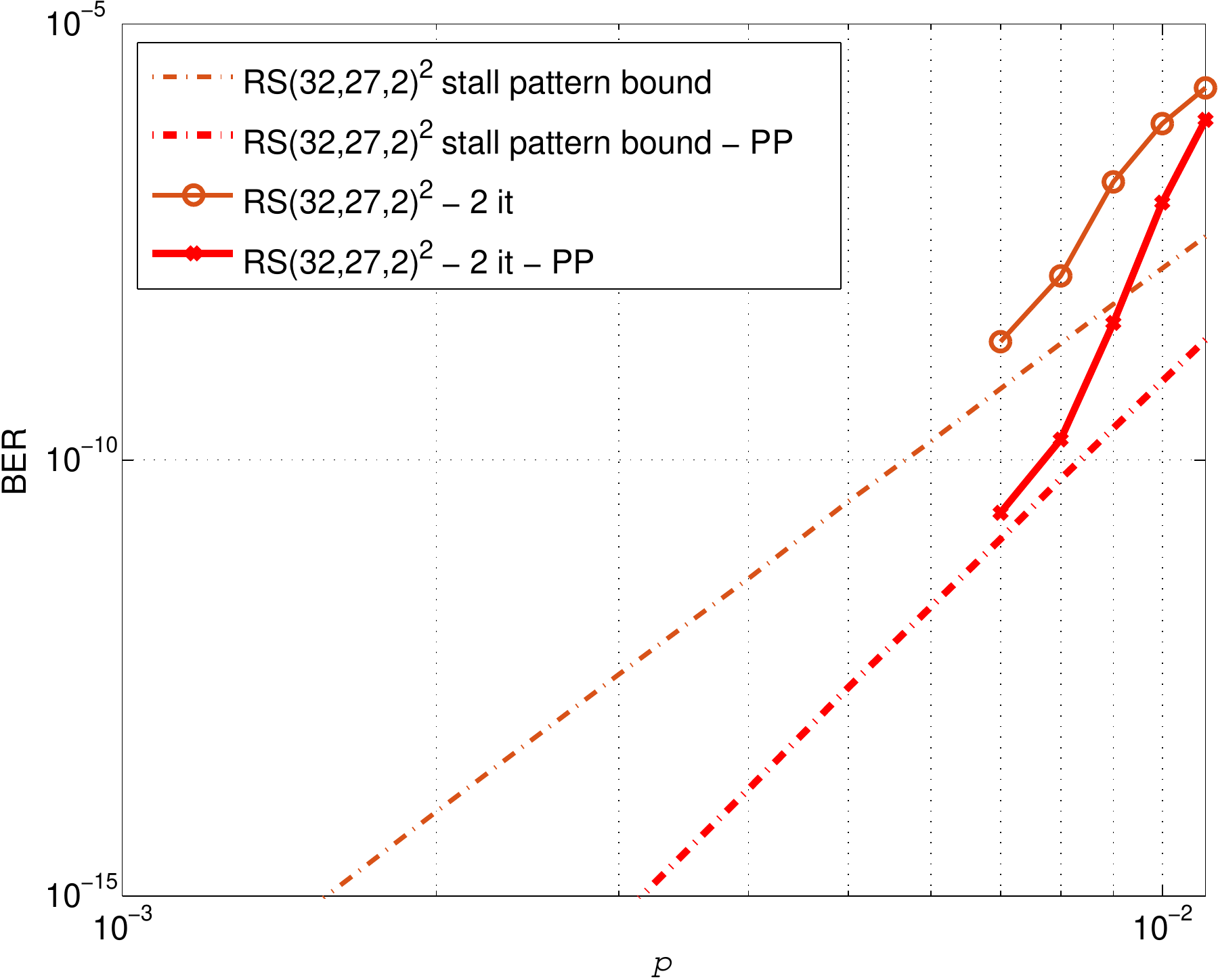}
    \caption{Stall-pattern contribution bounds and BER curves for extended RS-based (32,27,2)$^2$ product code.}
    \label{fig:RSbounds}
	\end{center}
\end{figure}

Similar results are shown in Figure~\ref{fig:RSbounds} for the (32,~27,~2) extended RS-based product code, with $b=5$. Two decoding iterations, with and without post processing, are sufficient to show how the BER curves closely follow the error bound. Post processing improves the error correction performance and decreases the stall-pattern contribution to the error floor by about two orders of magnitude.

\section{Conclusions}
\label{sec:conc}
In this work we have proposed a novel post-processing technique for product codes based on extended-polynomial component codes. Post-processing uses the knowledge of failed row- and column-component-code decoding to flip the bits at their intersection. This technique was shown to be able to both greatly lower the product-code error bound and to increase its steepness: on an example code, the error floor is lowered by more than three orders of magnitude at a channel BER of $2 \times 10^{-3}$. Simulation results show a comparable error correction performance improvement under various decoding conditions.

\bibliographystyle{IEEEtran}	

\end{document}

